\documentclass[a4paper]{jpconf}
\usepackage{graphicx}
\begin{document}
\title{Excitation function Analysis for Charmonium Production in Heavy-Ion Collisions}

\author{Kai Zhou}

\address{Institute for Theoretical Physics, Johann Wolfgang Goethe-University, Max-von-Laue-Str. 1, 60438 Frankfurt am Main, Germany}

\ead{zhou@th.physik.uni-frankfurt.de}

\begin{abstract}
Both color screening and regeneration are hot medium effects on charmonium production in heavy ion collisions. While they affect in an opposite way the charmonium
yield, their competition in transverse dynamics bring sensitivity to the ratio of averaged transverse momentum suqare for charmonium, which thus can reveal more
nature of the QCD medium created from the collisions. We make an excitation analysis based on transport approach to illustrate such a picuture. 
\end{abstract}

Since 1986 and firstly in light of color screening arguement for the initial proposing~\cite{satz86}, $J/\Psi$ has long been considered as an effective probe to
the hot and dense QCD matter created from heavy ion collisions. A suppression for their production in A+A compared with in p+p collisions
is monivated and extensively studied both in experiment and theory. Along with more released experimental data, it's
realized that apart from color screening, many other medium effects come into play during the collision evolution, which make
the picture for charmonium production complicated. Before any hot medium effects acting, the cold nuclear matter also
impact the $J/\Psi$ production through nuclear absorption, Cronin effect and nuclear shadowing effect. After entering the created
partonic medium, the color screening will alter the binding inside the bound state. Different suppression mechanism availably work
here at different temperature domain. When the temperature is higher than the charmonium dissociation temperature, no bound states survives based
on spectral analysis where no peak signal left due to color screening, thus they will melt into the medium.
Below this melting threshold, the partons inside the medium can break-up the quarkonium bound states through collisions, like the gluon dissociation and
quasi-free dissociation, which respectively correspond to Singlet-to-Octet transition and Landau damping in terms of EFT (effective field theory). Another important
concept arise here is regeneration~\cite{reg}, thus the recombination to bound state from those uncorrelated charm and anti-charm quark pairs
being in favored of their more abundant hard production along with increasing collision energy. Through regeneration, the hot medium effects on heavy
quarks would also leave a mark on heavy quarkonium production.

Actually from the theoretical point of view it's still not clear enough about the quarkonium evolution inside QGP. Like, which suppression mechnism
works dominantly? Which is the proper potential, free energy F or internal energy U, for quarkonium in medium? Also, in a rapidly cooling QGP medium,
a potential analysis might not possess enough validity since a bound state would hardly always lie in a fixed eigenstate of the evolving Hamiltonian describing in
medium quarkonium. Even further, if a seperate treatment for heavy quarkonia evolution taking the QGP
medium only as background really works is somehow questionable, see recently that has been pointed out in Ref.~\cite{kharzeev}, according to the associated large entropy for heavy
quarkonium in medium observed from Lattice QCD study (especially near $T_c$)~\cite{lat}, the coupling between medium constituent and quarkonium interquarks is stronger
even than the binding between the interquarks inside the bound state. This makes it unjustifiable to treat the bound state as an color neutral entity to be
influenced by the surrounding medium, since actually each of their interquarks are strongly entangled with the medium to be consistent with the Lattice results~\cite{lat}.

Nevertheless, in the present proceeding we provide a transport model based analysis for the medium effects on charmonium production, and we especially focus on
their excitation function to illustrate the picture given by kinetic approach. We now start with a brief introduction to the transport model we adopted in the
analysis. Since a quarkonium is heavy and can hardly reach equilibrium with the surronding medium, its phase space distribution function, $f_{\Psi}({\bf x},{\bf p},t)$,
is governed by the following Boltzmann-type transport equation~\cite{tsinghua},
\begin{equation}
{\partial f_\Psi\over \partial t} +{\bf v}_\Psi\cdot{\bf \nabla}f_\Psi=-\alpha_\Psi f_\Psi +\beta_\Psi,
\label{trans}
\end{equation}
where the loss and gain terms $\alpha$ and $\beta$ are the main hot medium effects describing dissociation and regeneration,
while the cold matter effects are reflected in the initial condition. Taking gluon dissociation as the dominant suppression mechanism,
$\alpha$ is the momentum integration of the dissociation cross section
multiplied by the thermal gluon distribution. The color screening induced melting is also considered, where the melting temperature $T_d$ is evaluated from potential
approach. Using detailed balance the regeneration cross section can be got from the gluon dissociation cross section. By converluting it with the in-medium charm quarks's
distribution we obtain the continuous regeneration rate $\beta({\bf x},{\bf p}_{\Psi},t)$. Considering the experimentally observed large quench factor~\cite{dquench} and
anisotropic flow~\cite{dflow} of open charm mesons, we here approximate the charm quarks' distribution to be kinetically thermalizaed which is a strong interaction limit.
On the level of kinetic approach, the medium information is then encoded in the loss and gain term through the distribution function, thus the appeared local temperature
$T({\bf x},t)$ and fluid velocity $u_{\mu}({\bf x},t)$ those being controled by ideal hydrodynamics in the present model analysis.
Using the above transport model, we have studied the nuclear modification factor $R_{AA}$ and transverse momentum distribution and also the anisotropic flow for charmonium
production~\cite{tsinghua}. Along with increasing collision energy, the medium becomes hotter and we naturally expect stronger $J/\Psi$ suppression to be observed.
However, it's found that in mid-rapidity the suppression magnitude of $J/\Psi$ at RHIC is similar to that at SPS and smaller than that at LHC. To make the advantages protruded in
explaining these phenomenon from transport model, here we turn to analyze their collision energy dependence thus the excitation function.

\begin{figure}[htbp]
\includegraphics[width=0.45\textwidth]{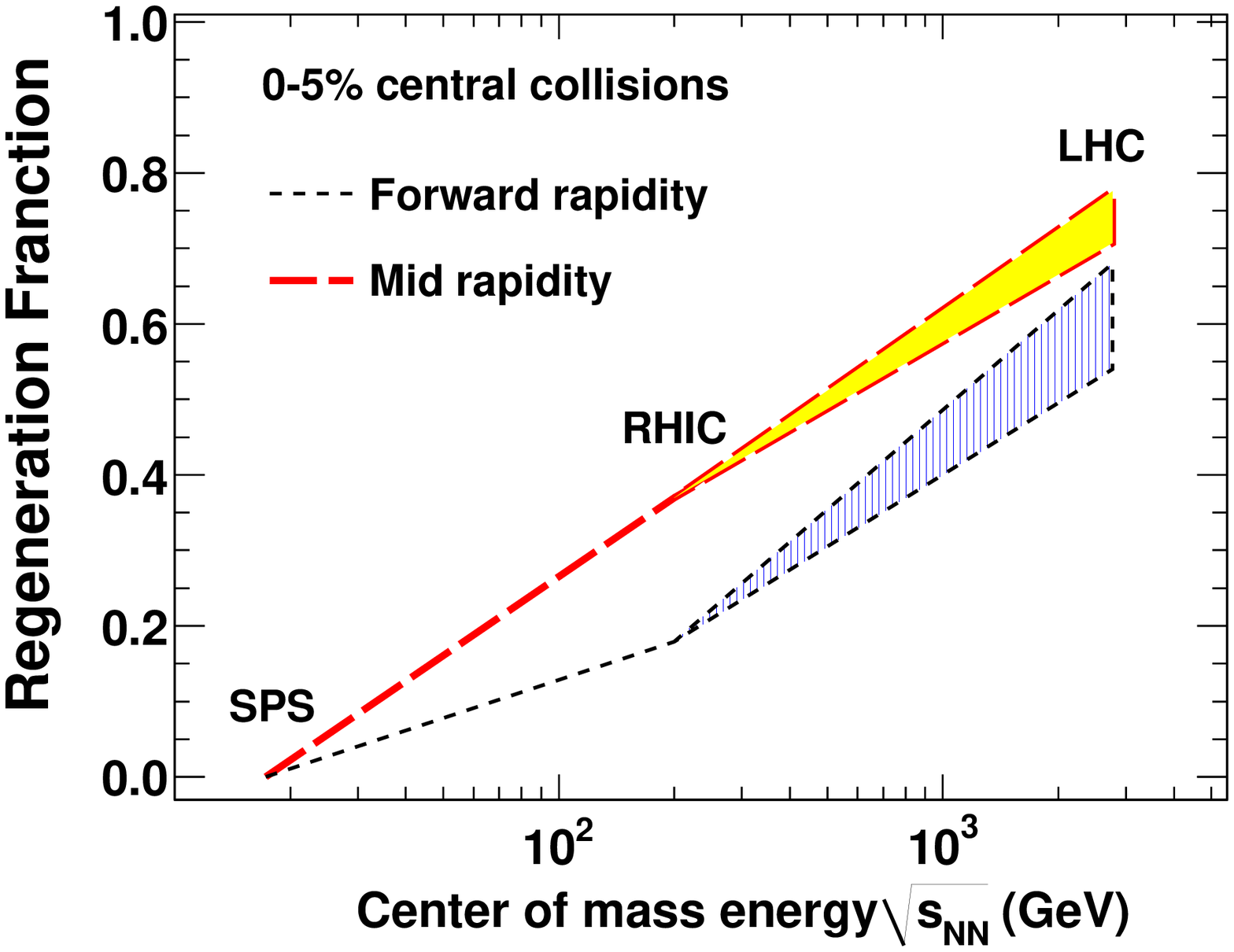}
\includegraphics[width=0.45\textwidth]{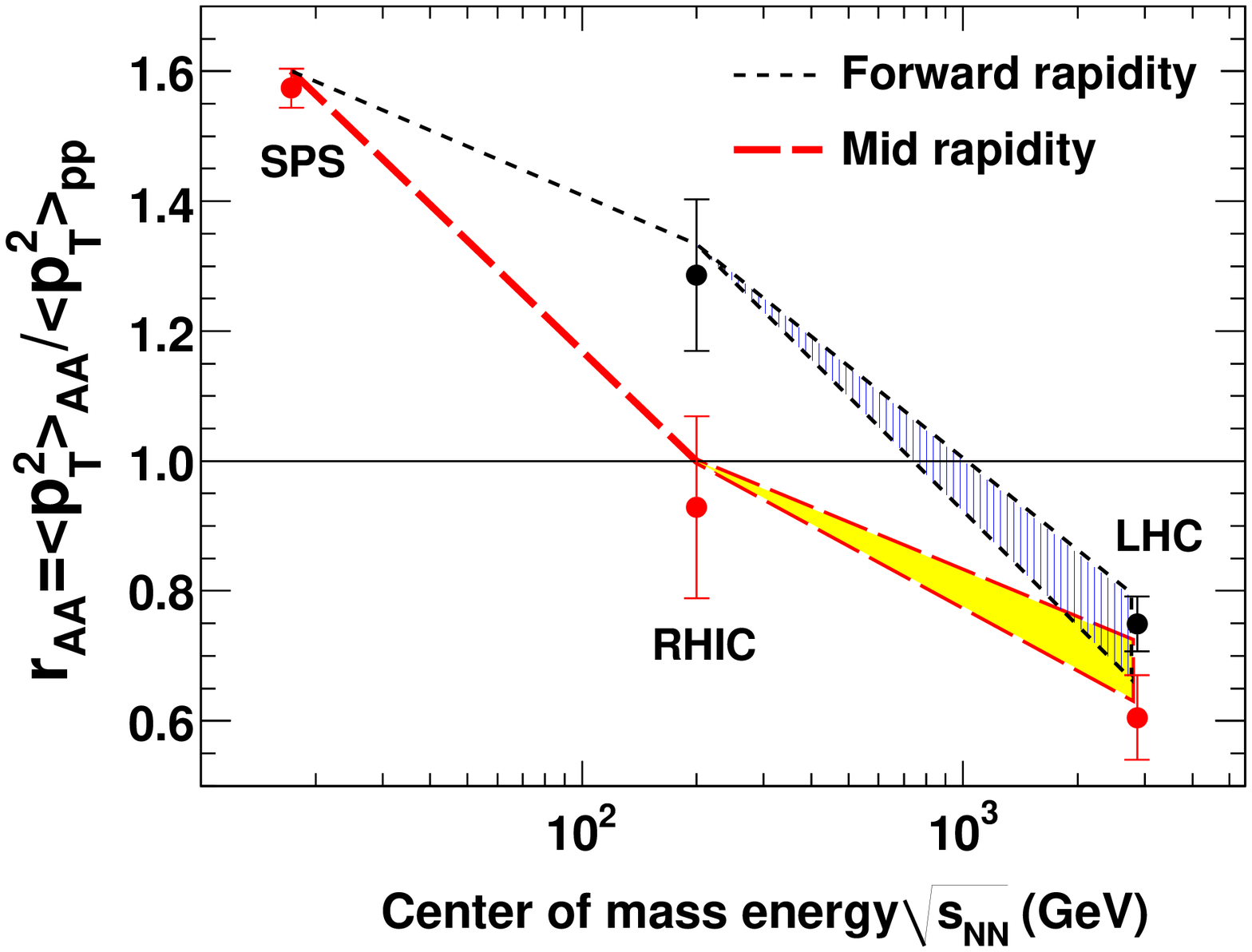}
\caption{(Color online) (left) $J/\psi$ regeneration fraction in
central Au+Au collisions at RHIC and Pb+Pb collisions at SPS and LHC.
(right) $r_{AA}=\langle p_t^2 \rangle_{AA}/\langle p_t^2\rangle_{pp}$
in central Au+Au collisions at RHIC and Pb+Pb collisions at SPS and LHC. The band at LHC comes from charm quark cross section's uncertainty.
Experimental data are also shown for SPS~\cite{sps}, RHIC~\cite{rhic} and LHC~\cite{lhc}.}
\label{fig1}
\end{figure}

As the collision becomes more violent, more charm quark pairs are produced through hard process, the regeneration componant gradually dominate the charmonium production. The fraction of the
regenerated $J/\Psi$s, $g_{AA}=N^{reg}_{AA}/N_{AA}$, calculated from our transport model, is shown in the left of Fig. \ref{fig1} as a function of collision
energy, where $N^{reg}_{AA}$ is $J/\Psi$ yield from regeneration. At lower energy collisions like at SPS, only very few charm quarks are produced
initially, then almost no regeneration happenes. At RHIC, the initial production is still dominant but the regeneration becomes equally important especially in mid-rapidity.
At LHC, the final charmonium yield is dominated by regeneration in both mid and forward rapidiy. The lower fraction in forward rapidity merely reflects the rapidity
distribution of charm quarks.

Since the transverse motion in heavy ion collisions which is developed during the dynamical evolution of the system can directly reflect the multiple scattering and also
parton density in the medium, as has been well documented in light quarks sector~\cite{light}, for charmonium production, the transverse momentum distribution also can
reveal more information about their evolution and be more sensitive to the medium effects. We propose the following new defined nuclear modification factor for transverse
momentum square
\begin{equation}
r_{AA}=\frac{\langle p_t^2 \rangle_{AA}}{\langle p_t^2\rangle_{pp}}.
\end{equation}
We show in the right of Fig. \ref{fig1} the energy dependence of $r_{AA}$ in central collisions. As been shown in left of Fig. \ref{fig1},
at lower energy collisions almost all of the finally observed charmonia are from survived previously produced ones, then the initial state
gluon multiple scattering, i.e. the Cronin effect, is dominant and together with leakage effect and medium suppression would increase their transverse momentum leading to $r_{AA}>1$. Contrarily, at very high energy collisions,
LHC, due to the copiourly production of charm quarks, the recombination of c and $\bar c$ into charmonium, thus the regeneration becomes voilently significant and
dominate the whole charmonium production finally. Since the heavy quark would be quenched when passing through the medium, the regenerated $J/\Psi$s sourced from the
in-medium charm quarks thus possess lower transverse momentum compared to initial prpduction. In between the above two limits, at RHIC, in mid-rapidity
the competition between initial production which controls high $p_T$ charmonium production and the regeneration which inherits the low momentum of kinetically thermalized
charm quarks leads to a cancelled influence on transver momentum square, while in forward rapidity $r_{AA}$ is around 1.3 due to the still overwhelming initial production.
Our prediction is also confirmed by the experimental measurements as shown in the plot. Being unlike the yield $R_{AA}$ which appears cancelling out each others increasing
between regeneration and suppression for initial production, the newly defined $r_{AA}$ lends additional sensitivity to the transverse dynamics and present a robust
excitation feature: as a consequence of the increasing regeneration with collision energy, the interplay between regeneration and initial contribution just leads to a
decreasing trend for $r_{AA}$ from SPS to LHC, manifesting a hotter medium creating and stronger hot medium effects impacting in higher energy collisions.

\begin{figure}[t]
  \centering
    \begin{minipage}[c]{.6\textwidth}
    \centering
    \includegraphics[width=0.8\textwidth]{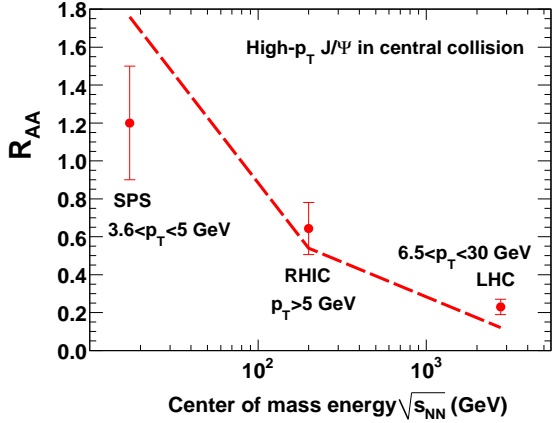}
  \end{minipage}
  \hspace{-0.2cm}
  \begin{minipage}[c]{.35\textwidth}
    \centering
    \caption{High $p_T$ $J/\psi$s' $R_{AA}$ in mid-rapidity in
central Au+Au collisions at RHIC and Pb+Pb collisions at SPS and LHC.
Experimental data are shown for SPS~\cite{sps2}, RHIC~\cite{rhic2} and LHC~\cite{lhc2}.}
    \label{fig2}
  \end{minipage}%
\end{figure}

From another aspect, due to the strong interaction with the medium, the charm quarks' distribution becomes steeper although they are initially hard, consequently
the regeneration for charmonium
decrease with increasing momentum and hardly contribute to high $p_T$ $J/\Psi$s. Therefore, the high $p_T$ $J/\Psi$s are characterized solely by Debye screening
effect and also medium induced suppression on the initially produced charmonia. Fig. \ref{fig2} shows the excitation function for high $p_T$ $J/\Psi$s' $R_{AA}$ in central collisions. The decreasing
trend obviously re-authenticate the natural expectation : stronger hot medium effects are active in higher energy collisions.

To summarize, we analyze the excitation behavior of charmonium production in heavy ion collisions based on transport model. The color screening and regeneration are both
medium induced effects on charmonium production, they affect the production yield in an opposite way. At the same time, they drastically influence the transverse momentum
dynamics for charmonium. We promote the ratio $r_{AA}=\langle p_t^2 \rangle_{AA}/\langle p_t^2\rangle_{pp}$ for charmonia as a much more sensitive probe to the
hot and dense QCD medium, its decresing trend along with collision energy robustly indicate a hotter QCD matter created from higher energy collsions. The high $p_T$ $J/\Psi$
from another angle also helps to distringuish the medium among different energy collisions.

\section*{Acknowledgments}
\quad I'd like to acknowledge financial support from Frankfurt Institute for Advanced Studies.

\end{document}